\documentclass[iop]{emulateapj}

\usepackage{graphicx}
\usepackage{amssymb}
\usepackage{epstopdf}
\usepackage{natbib}
\usepackage{multirow}
\usepackage{hhline}
\usepackage{verbatim}
\usepackage[normalem]{ulem}
\usepackage[backref,breaklinks,colorlinks,citecolor=blue]{hyperref}
\usepackage[all]{hypcap}

\shorttitle{MAIN SEQUENCE}
\shortauthors{LEE ET AL.}

\newcommand{\lir}{L_{\rm{IR}}}

\begin{document}

\title{A Turnover in the Galaxy Main Sequence of Star Formation at $M_{*} \sim 10^{10} M_{\odot}$ for Redshifts $z < 1.3$}
\author{Nicholas Lee\altaffilmark{1}, D. B. Sanders\altaffilmark{1}, Caitlin M. Casey\altaffilmark{1,2}, Sune Toft\altaffilmark{3}, N.Z. Scoville\altaffilmark{4}, Chao-Ling Hung\altaffilmark{1,5}, Emeric Le Floc'h\altaffilmark{6}, Olivier Ilbert\altaffilmark{7}, H. Jabran Zahid \altaffilmark{1,5}, Herv\'{e} Aussel\altaffilmark{6}, Peter Capak\altaffilmark{4,8}, Jeyhan S. Kartaltepe\altaffilmark{9,10}, Lisa J. Kewley\altaffilmark{11}, Yanxia Li\altaffilmark{1}, Kevin Schawinski\altaffilmark{12}, Kartik Sheth\altaffilmark{13}, Quanbao Xiao\altaffilmark{1,14}}
%\affil{Institute for Astronomy}
%\affil{2680 Woodlawn Dr., Honolulu, HI, 96822}
\altaffiltext{1}{Institute for Astronomy, 2680 Woodlawn Dr., Honolulu, HI, 96822, USA}
\altaffiltext{2}{Department of Physics and Astronomy, University of California, Irvine, CA, 92697, USA}
\altaffiltext{3}{Dark Cosmology Centre, Niels Bohr Institute, University of Copenhagen, Juliana Mariesvej 30, 2100 Copenhagen, Denmark}
\altaffiltext{4}{California Institute of Technology, MS 105-24, 1200 East California Boulevard, Pasadena, CA 91125}
\altaffiltext{5}{Harvard-Smithsonian Center for Astrophysics, Cambridge, MA 02138, USA}
\altaffiltext{6}{UMR AIM (CEA-UP7-CNRS), CEA-Saclay, Orme des Merisiers, b\^{a}t. 709, F-91191 Gif-sur-Yvette Cedex, France}
\altaffiltext{7}{Aix Marseille Universit\'e, CNRS, LAM (Laboratoire d'Astrophysique de Marseille) UMR 7326, 13388, Marseille}
\altaffiltext{8}{Spitzer Science Center, 314-6 Caltech, 1201 East California Boulevard, Pasadena, CA 91125}
\altaffiltext{9}{Hubble Fellow}
\altaffiltext{10}{National Optical Astronomy Observatory, 950 North Cherry Ave., Tucson, AZ, 85719, USA}
\altaffiltext{11}{Research School of Astrophysics, Australian National University, Canberra ACT 0200, Australia}
\altaffiltext{12}{Institute for Astronomy, Department of Physics, ETH Zurich, Wolfgang-Pauli-Strasse 27, CH-8093 Zurich, Switzerland}
\altaffiltext{13}{National Radio Astronomy Observatory, 520 Edgemont Road, Charlottesville, VA 22903, USA}
\altaffiltext{14}{The Shanghai Key Lab for Astrophysics, 100 Guilin Road, Shanghai, 200234, People's Republic of China}

\begin{abstract}

The relationship between galaxy star formation rates (SFR) and stellar masses ($M_\ast$) is re-examined using a mass-selected sample of  $\sim$62,000 star-forming galaxies at $z \le 1.3$ in the COSMOS 2-deg$^2$ field.   Using new far-infrared photometry from $Herschel$-PACS and SPIRE and $Spitzer$-MIPS 24 $\mu$m, along with derived infrared luminosities from the NRK method based on galaxies' locations in the restframe color-color diagram $(NUV - r)$ vs. $(r - K)$, we are able to more accurately determine total SFRs for our complete sample.  At all redshifts, the relationship between median $SFR$ and $M_\ast$ follows a power-law at low stellar masses, and flattens to nearly constant SFR at high stellar masses.  We describe a new parameterization that provides the best fit to the main sequence and characterizes the low mass power-law slope, turnover mass, and overall scaling. The turnover in the main sequence occurs at a characteristic mass of about $M_{0} \sim 10^{10} M_{\odot}$ at all redshifts. The low mass power-law slope ranges from 0.9-1.3 and the overall scaling rises in SFR as a function of $(1+z)^{4.12 \pm 0.10}$. A broken power-law fit below and above the turnover mass gives relationships of $SFR \propto M_{*}^{0.88 \pm 0.06}$ below the turnover mass and $SFR \propto M_{*}^{0.27 \pm 0.04}$ above the turnover mass.  Galaxies more massive than $M_\ast \gtrsim 10^{10}\ M_{\rm \odot}$ have on average, a much lower specific star formation rate (sSFR) than would be expected by simply extrapolating the traditional linear fit to the main sequence found for less massive galaxies.

\begin{comment}
We study the relationship between star formation rate (SFR) and stellar mass ($M_{*}$) in a mass-selected sample of star-forming galaxies in the COSMOS field.  Using recent results from {\em Herschel} PACS and SPIRE, we are able to accurately determine the amount of star-formation that is obscured by dust. After testing the accuracy of various star-formation rate indicators from short wavelength data (in the UV \& optical), we construct a ladder of star-formation rate indicators and accurately measure the SFR of a mass-selected sample of star-forming galaxies from $0.2 < z < 1.3$.  At all redshifts, the relationship between SFR and $M_{*}$ is linear at low masses, but turns over and begins to flatten at stellar masses above $M_{*} > 10^{10} M_{\odot}$, such that the trend is better fit by a 2nd order polynomial than the single linear main sequence relationship often discussed in the literature. On average, galaxies more massive than $M_{*} \sim 10^{10}$ have a lower specific star formation rate than would be expected from a purely linear main-sequence, which may suggest that many of these galaxies are undergoing some sort of quenching. In addition, high mass galaxies have a much higher fraction of ``starburst'' galaxies, with either significantly elevated SFRs or extremely red colors, both of which could be caused by major mergers. 
\end{comment}
\end{abstract}

\keywords{galaxies: evolution - galaxies: high-redshift }

\section{Introduction}

Over the last decade, a tight correlation between a galaxy's star formation rate ($SFR$) and its stellar mass ($M_{*}$) has been discovered \citep{2007ApJ...660L..43N,2007ApJ...670..156D,2007A&A...468...33E}.  Commonly referred to as the galaxy ``main-sequence'' (MS) of star formation, this relationship has important implications for the physical nature of star formation in galaxies. The MS is generally described as a single power law of the form $SFR \propto M_{*}^{\beta}$, with $\beta = 0.7$--1.0 and the normalization of the MS evolving to higher values at increasing redshift \citep{2007ApJ...660L..43N}.  

A common interpretation of the existence and tightness of the main-sequence is that the majority of star-forming galaxies are powered by similar quasi-steady processes, with only a small fraction of galaxies undergoing more chaotic processes such as major merger events that might be expected to produce strong bursts of star formation  \citep[e.g.][]{2011A&A...533A.119E,2011ApJ...739L..40R,2012ApJ...747L..31S}.  These starburst galaxies are generally thought to lie significantly above the MS and represent a minority of galaxies.

A key uncertainty in measuring galaxy SFRs is the effect of dust obscuration. The most direct method of determining dust obscuration is from observations in the far-infrared, where the absorbed starlight is thermally reradiated. In the absence of far-infrared data, various extrapolations from shorter wavelength have been used to study the main sequence, such as using emission lines combined with reddening corrections to infer the dust-corrected SFR \citep{2010ApJ...716..348B,2012MNRAS.420.1926S,2012ApJ...757...54Z,2013ApJ...777L...8K} or measuring the UV or optical emission from young massive stars and correcting for the radiation lost to dust obscuration \citep{2011ApJ...733...99L,2011ApJ...739L..40R,2014ApJ...791L..25S}. Observations in the mid-infrared (e.g. 24 $\mu$m) have been used to estimate far-infrared luminosities \citep[e.g.][]{2007ApJ...660L..43N,2007ApJ...670..156D,2007A&A...468...33E}, although the accuracy of these estimates decreases at high redshifts and bright infrared luminosities \citep[e.g.][]{Papovich:2007p38,2010ApJ...717..175L,2011A&A...533A.119E}. Studies of radio emission take advantage of the well-known radio-FIR correlation \citep{1985ApJ...298L...7H,1992ARA&A..30..575C,2001ApJ...554..803Y} to estimate the infrared luminosities, but many of these studies rely on stacking to overcome high sensitivity limits \citep{2009MNRAS.394....3D,2009ApJ...698L.116P,2011ApJ...730...61K}.  The consensus from these studies is that the MS follows a single power law $SFR \propto M_{*}^{\beta}$, with the slope generally between $\beta = 0.7$--1.0 and the normalization varying based on the study's redshift, SFR indicator, sample selection, and IMF \citep[for a summary, see][]{2014arXiv1405.2041S}. 

However, a few studies have found indications of a more complex main-sequence relationship. Some studies suggest that the MS slope varies with stellar mass so that a single power-law cannot explain the MS and a stellar mass-dependent slope is a better fit \citep{2011ApJ...730...61K,2012ApJ...754L..29W,2014A&A...561A..86M}. Recent studies based on far-infrared selected samples from {\em Herschel} show that far-infrared selected galaxies lie mostly above the MS, with a much shallower slope in ${\rm log}(SFR)/{\rm log}(M_{*})$ \citep{2013ApJ...778..131L,2013A&A...554L...3O,2013MNRAS.435..158O,2013arXiv1311.5228L}. However, this discrepancy is due to the flux limited selection of far-infrared samples that introduce a SFR-based selection bias, as compared to studies based on stellar mass-selected galaxy samples. This has been demonstrated by stacking analyses that explore the far-infrared emission as a function of stellar mass and find generally good agreement between dust-corrected UV-derived SFRs and {\em Herschel}-derived SFRs \citep[e.g.][]{2014MNRAS.443...19R}.

Stacking is a commonly used technique to measure low-level emission from galaxies that would be undetected individually.  Stacking analyses require a number of assumptions and can miss vital information about individual galaxies and their distributions. Unless the parent population is identical (a key assumption in stacking), interpretation of stacking results can be difficult because the underlying distribution is unknown \citep[although see][for a possible method to determine the underlying distribution]{2014arXiv1409.5433S}. In addition, these stacking analyses do not explain why the dust-corrected SFRs cannot accurately recover the SFRs seen in high luminosity galaxies, which have an elevated contribution to the integrated build up of stellar mass in the universe. 

Direct {\em Herschel} FIR measurements remain a unique tool to properly estimate the ongoing star formation rate in the most active dusty galaxies. Analysis of the rest-frame UV emission in dusty galaxies suggests that applying the nominal attenuation laws \citep[e.g.][]{1999ApJ...521...64M,2000ApJ...533..682C} will dramatically underestimate total star formation rate in galaxies exceeding $\sim 50 M_{\odot}/$yr  \citep{2004ApJ...616...71S,2014ApJ...796...95C,2014MNRAS.443...19R}.

In the following paper, we attempt to address these issues by using a dust-corrected SFR indicator that is accurate for galaxies at all luminosities. By analyzing a large sample of individual galaxies, we do not lose information about the distribution of sources from stacking and can re-examine the shape of the star-forming MS in a stellar mass-selected sample. The data are described in \autoref{sec:data} and SFRs computed by several different methods are measured and compared in \autoref{sec:sfr}.  In \autoref{sec:ms} we analyze our mass-selected sample of galaxies in the $SFR/M_{*}$ plane and find the best fits to the data. The implications of the main-sequence are discussed in \autoref{sec:discussion} and we list our conclusions in \autoref{sec:conclusions}.  When calculating rest-frame quantities, we use a cosmology with $\Omega_{m} = 0.28$, $\Lambda = 0.72$, and $H_{0} = 70$ km s$^{-1}$ Mpc$^{-1}$ \citep{2013ApJS..208...19H}.  A \citet{2003PASP..115..763C} Initial Mass Function (IMF) truncated at 0.1 and 100 $M_{\odot}$ is used when deriving SFRs and stellar masses.

\section{Data}\label{sec:data}

Our analysis of the MS is made possible by the large area multi-wavelength coverage of the COSMOS field, a 2 deg$^{2}$ area of the sky with observations from the ultraviolet through far-infrared and radio \citep{2007ApJS..172....1S}. We construct a mass-complete sample of galaxies with $K_{s} < 24$ from the deep $K_{s}$-band catalog of \citet{2013A&A...556A..55I}, based on data from the first UltraVISTA DR1 data release, covering $\sim$75\% of the COSMOS field \citep{2012A&A...544A.156M}.  20 bands of optical and near-infrared photometry were extracted using matched apertures in dual-image mode from the various available COSMOS images and was combined with GALEX magnitudes from the multi-wavelength catalog of \citet{2007ApJS..172...99C}.  We use updated {\em Spitzer} IRAC photometry from the Spitzer Large Area Survey with Hyper-Suprime-CAM (SPLASH, Capak et al. in prep).  We cross-match this catalog with the {\em Spitzer} MIPS 24 $\mu$m catalog of \citet{2009ApJ...703..222L} and the {\em Herschel} catalog of \citet{2013ApJ...778..131L} using a matching radius of $2''$.  

\subsection{Source Selection}

We interpolate the 90\% stellar mass completeness limits from \citet{2013A&A...556A..55I} to determine approximate mass completeness thresholds at all redshifts, and select only galaxies with stellar masses above their redshift dependent mass completeness limit. When studying star-forming galaxy populations, we separate ``star-forming'' and ``quiescent'' galaxies using a two-color selection technique: NUV$ - r^{+}$ versus $r^{+} - J$ \citep[as described in][]{2013A&A...556A..55I}.  Specifically, galaxies with absolute magnitude colors $M_{\rm{NUV}} - M_{r} > 3(M_{r} - M_{J}) + 1$ and $M_{NUV} - M_{r} > 3.1$ are considered ``quiescent'' ($\sim 15$\% of the sample) while the remaining galaxies are considered actively star-forming galaxies. 

Star-forming galaxies that also contain luminous AGN are a concern because the luminosity from the AGN is extremely difficult to separate from emission from star-formation, and thus these sources may have erroneously high SFRs (although this concern is lessened for FIR sources because AGN generally heat dust to temperatures too hot to radiate in the far-infrared).  On the other hand, many of the galaxies that host AGN also contain significant star-formation, and removing these sources introduces a bias to our study. We find that the overall results of our study are not significantly affected by either the inclusion or exclusion of these sources, so we do not remove galaxies that have been detected in the X-ray ($\sim 0.5$\% of sample) by {\em XMM-Newton} \citep{2010ApJ...716..348B} or {\em Chandra} \citep{2011ApJ...741...91C}, or that have IRAC power-law colors \citep{2008ApJ...687..111D} that suggest AGN activity ($\sim 25$\% of sample). 

\subsection{Infrared Data}\label{sec:ir_data}

The {\em Herschel}-selected sample of galaxies is described in detail in \citet[][hereafter L13]{2013ApJ...778..131L} and is briefly summarized here.  L13 use {\em Spitzer} 24 $\mu$m and VLA 1.4 GHz priors to find 4,218 sources in COSMOS that were each detected in at least two of the five available {\em Herschel} PACS (100 $\mu$m or 160 $\mu$m) and SPIRE (250 $\mu$m, 350 $\mu$m, or 500 $\mu$m) bands.  These sources span log$(\lir/L_{\odot}) = 9.4$--13.6 and $z = 0.02$--3.54.  Dust properties of each source (e.g. $\lir$, $T_{\rm{dust}}$, $M_{\rm{dust}}$) were measured by fitting the full infrared photometry to a coupled modified blackbody plus mid-infrared power law using the prescription given in \citet{2012MNRAS.425.3094C} and assuming an opacity model where $\tau = 1$ at 200 $\mu$m.  

There is a population of galaxies that are classified as ``quiescent'' from their NUV$ - r^{+}$ versus $r^{+} - J$ colors, but have been detected in the infrared by {\em Herschel} or {\em Spitzer}, suggesting that these galaxies are actually undergoing a significant amount of star-formation ($\sim 7$\% of the ``quiescent'' population).  These galaxies are more consistent with being very dusty objects that have extremely red colors due to obscuration, not lack of active star-formation, so we include these galaxies in our sample of star-forming galaxies (see \autoref{sec:quiescent}). 

\subsection{Photometric Redshifts and Physical Parameters}

\citet{2013A&A...556A..55I} measure accurate 30-band photometric redshifts of the full $K_{s}$-band COSMOS catalog. We find a median $\Delta z/(1+z) = 0.02$ in our sample of star-forming galaxies, with a catastrophic failure ($|\Delta z|/(1+z) > 0.15$) in 5.6\% of sources. In addition, physical parameters such as stellar mass and star formation rate have been calculated by fitting the Spectral Energy Distributions (SEDs) to synthetic spectra generated using the Stellar Population Synthesis (SPS) models of \citet{2003MNRAS.344.1000B}. We also recalculate the physical parameters using different templates and extraction parameters, and find that our final results are not affected by the specific choice of template. Thus, we use the same set of parameters used to create the catalog in \citet{2013A&A...556A..55I}, but with updated near-IR photometry from SPLASH. 

\section{Analysis}\label{sec:sfr}

\subsection{Star Formation Rate Calculations}

There are many methods for estimating a galaxy's SFR based on observations at various wavelengths \citep[for a review, see][]{1998ARA&A..36..189K,2011ApJ...737...67M}. Here we compare a few commonly used SFR indicators using a subset of COSMOS galaxies to determine how much the different SFR methods disagree. In all cases, we measure the total SFR as $SFR_{\rm{Tot}} = SFR_{\rm{IR}} + SFR_{\rm{UV}}$.  

\subsubsection{Infrared derived $SFR$}

As discussed in \autoref{sec:ir_data}, the infrared properties of the {\em Herschel}-selected galaxies have been measured by fitting the infrared SEDs to a coupled modified blackbody plus mid-infrared power law model \citep{2012MNRAS.425.3094C}.  It has been shown that measuring the $\lir$ from fitting the far-infrared data to libraries of SED \citep[e.g.][]{Chary:2001p1425,2002ApJ...576..159D} gives roughly the same results as the modified blackbody plus power-law model \citep{2012MNRAS.425.3094C,2012ApJS..203....9U,2013ApJ...778..131L}. The infrared observations give us an estimate of the obscured SFR, and we combine this with UV observations of the unobscured SFR to derive the total SFR as in \citet{2013A&A...558A..67A}:
\begin{equation} 
SFR_{\rm{Total}} = (8.6 \times 10^{-11}) \times (\lir + 2.3 \times \nu \kern-0.4ex L_{\nu}(2300 \rm{\AA})) \label{eqn:sfr}
\end{equation}
where $L_{\rm{IR}} \equiv L(8$--$1000 \mu$m) and all luminosities are measured in units of $L_{\odot}$. For sources with $SFR \gtrsim 50 M_{\odot}$/yr, the infrared contribution dominates the total SFR, contributing as much as $\sim 90$\% of the total SFR. 

While we have excellent {\em Herschel} coverage of the full 2-deg$^{2}$ COSMOS field that yields 4,218 sources, the detection limits of {\em Herschel} introduce a selection bias against all but the most luminous infrared sources.  A common method of determining the $L_{\rm{IR}}$ of less luminous galaxies is to use deep {\em Spitzer} 24 $\mu$m data to estimate the far-infrared luminosity \citep[e.g.][]{2009ApJ...703.1672K,2009ApJ...692..556R,2013ApJ...767...73R}. COSMOS has extremely deep coverage at 24 $\mu$m and \citet{2009ApJ...703..222L} provide SFR estimates for 36,635 galaxies, which we use to extend our sample of infrared detected galaxies to more moderate luminosities. 

\subsubsection{Optical \& UV based SFR Indicators}\label{sec:opt_SFR}

For galaxies without direct measurements from far- or mid-infrared wavelengths of the obscured SFR, the amount of radiation obscured by dust must be estimated indirectly. A common method for estimating total SFR is to fit libraries of model SEDs (that include prescriptions for dust obscuration) to optical \& UV photometry. \citet{2013A&A...556A..55I} use the full optical COSMOS photometry and fit to a library of synthetic spectra from \citet{2003MNRAS.344.1000B}, and estimate the total SFR for each of the galaxies in our sample from the best fit SEDs.  

Another method of estimating dust-corrected SFRs is by using rest-frame UV observations to measure the unobscured SFR and inferring the appropriate dust correction factor from observed colors. Two examples of this are the BzK method from \citet{2004ApJ...617..746D} and the NRK method from \citet{2013A&A...558A..67A}.  BzK SFRs are determined by using the observed-frame $B$-band photometry to measure the rest-frame UV luminosity, and then estimating the extinction as $E(B-V) = 0.25(B - z + 0.1)_{AB}$ \citep{2007ApJ...670..156D}. BzK SFRs are only valid for redshifts $1.4 < z < 2.5$, as these are the only redshifts where the desired portions of the SED are redshifted to the correct wavelengths, and we limit our selection to the {\em good}-sBzK with errors $\delta$log[SFR(UV)]$< 0.3$ dex \citep{2014MNRAS.443...19R}. 

NRK SFRs are calculated by using their location in the rest-frame color-color diagram $(NUV - r)$ vs. $(r - K)$ to estimate extinction. \citet{2013A&A...558A..67A} find that at $z \le 1.3$, the infrared excess $IRX\equiv \lir/L_{NUV}$ in star-forming galaxies can be parameterized as a function of redshift and the vector {\bf {\em NRK}}$ = 0.31 \times (\rm{NUV} - r) + 0.95 \times (r - K)$. This allows us to estimate the $L_{\rm{IR}}$ and calculate total SFR using \autoref{eqn:sfr}. When measuring NRK SFRs, we use the small ``sSFR correction'' as described in \citet{2013A&A...558A..67A}. 

\subsection{Comparison of SFR indicators}\label{sec:sfr_comp}

We compare commonly used SFR indicators using a common subset of COSMOS galaxies to determine how much agreement there is between the different measures of SFR. We have a large set of {\em Herschel} detected galaxies from \citet{2013ApJ...778..131L} where, for the first time, we have direct measurements of both the obscured and unobscured SFR (from UV observations) at a wide range of redshifts.  We compare the other SFR indicators discussed previously (24 $\mu$m, SED fits, BzK, and NRK) to this sample of 4,218 {\em Herschel} detected galaxies.  

\begin{figure*}[!h]
\centering
\includegraphics[width=14 cm]{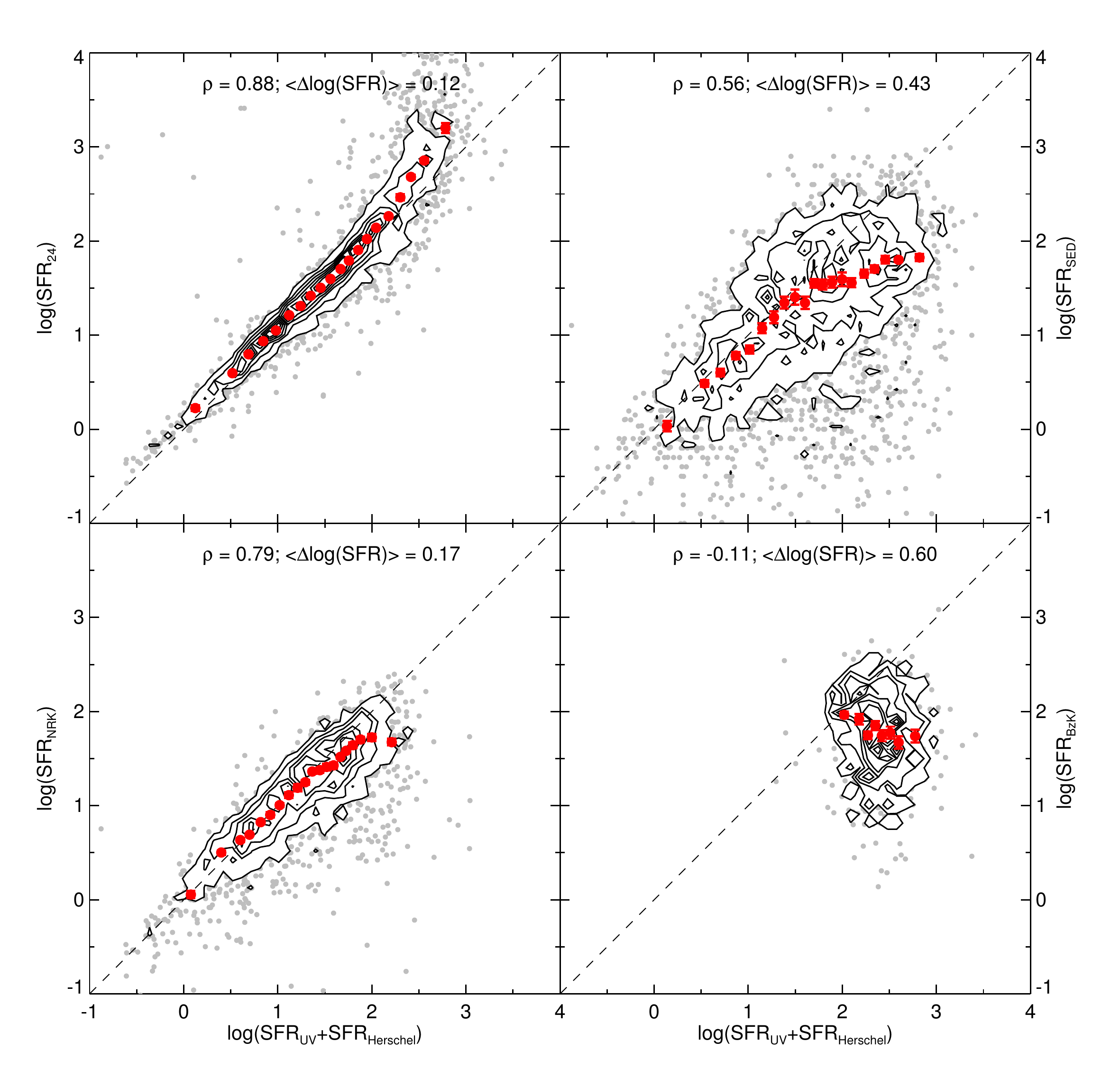}
\caption{Comparison of total SFR determined from combining direct measurements of FIR \citep[{\em Herschel},][]{2013ApJ...778..131L} and UV \citep[GALEX,][]{2007ApJS..172..468Z} with various SFR indicators using observations at shorter wavelengths. The different SFR indicators are: [top left] {\em Spitzer} 24 $\mu$m \citep{2009ApJ...703..222L} + GALEX; [top right] multi-wavelength SED fits \citep{2013A&A...556A..55I}; [bottom left] NRK \citep[$0 < z < 1.3$,][]{2013A&A...558A..67A}; and [bottom right] BzK \citep[$1.4 < z < 2.5$,][]{2007ApJ...670..156D}. In each panel, black contours give the density and concentration of sources, with extreme outliers plotted as gray circles. We bin the data in 20 equally populated bins (except BzK, which has 8 bins) and find the median $SFR_{indicator}$ in each bin. Errors on the median points are measured using a bootstrapping technique and are plotted when larger than the size of the symbol. At the top of each panel is the Pearson correlation coefficient (with a value of +1 indicating strong positive correlation and 0 indicating no correlation) and the typical difference between each particular SFR indicator and the {\em Herschel}-derived SFR.}
\label{fig:sfr_comp}
\end{figure*}

\autoref{fig:sfr_comp} displays the comparison of $SFR_{\rm{Total}}$ from the four different indicators discussed above to $SFR_{\rm{Total}}$ as measured by {\em Herschel}. Density contours show the location and concentration of the majority of the sources, with outliers shown in gray circles. Median values in 20 equally populated bins of $SFR_{\rm{Total,Herschel}}$ are over-plotted to show average trends. To determine the strength of the correlation between each SFR indicator and $SFR_{\rm{Total,Herschel}}$, we measure the Pearson correlation coefficient ($\rho$) and provide these values at the top of each sub-panel. The Pearson correlation coefficient can vary between +1 and -1, with +1 indicating total positive correlation, 0 indicating no correlation, and -1 indicating total negative correlation. We also measure the median difference between each SFR indicator and Herschel SFR ($<\Delta log(SFR)>$) and list these values at the top of each sub-panel. 

The 24 $\mu$m-determined SFR correlates with the {\em Herschel} SFR very well ($\rho_{24} = 0.88$, $<\Delta log(SFR_{24})>$ $=0.12$), except at the highest IR luminosities.  This trend has been previously explored in many studies which find that at moderate redshifts and IR luminosities, 24 $\mu$m observations are a good proxy for $\lir$, but at high redshifts and infrared luminosities, the 24 $\mu$m estimates tend to overpredict the true $\lir$, possibly due to redshifting of the observed 24 $\mu$m-band to wavelengths contaminated by PAH features \citep[e.g.][]{Papovich:2007p38,2010ApJ...717..175L,2011A&A...533A.119E}. As we are using the 24 $\mu$m SFRs to fill in the low and moderate luminosity galaxies that are not detected with {\em Herschel}, this discrepancy is not a major issue for our work.

The NRK SFRs also show strong correlation with the {\em Herschel}-derived SFRs ($\rho_{NRK} = 0.79$, $<\Delta log(SFR_{NRK})>$ $= 0.17$).  This is not completely unexpected since the NRK method was developed using 24 $\mu$m-derived SFRs as a baseline, but the NRK measured SFRs match very well with those derived from {\em Herschel}. Like with $SFR_{24}$, the correlation shows signs of breaking down at the highest SFRs, but as long as the NRK is used mainly for low SFR galaxies, it provides a reliable estimate of the SFR. 

By contrast, the agreement between $SFR_{\rm{SED}}$ and $SFR_{\rm{Total}}$ is quite poor, showing much weaker correlation between the two indicators ($\rho_{SED} = 0.56$). The tightness of the correlation is also much broader ($<\Delta log(SFR_{SED})>$ $=0.43$), even at low SFRs where the median points lie closer to the unity line. Again, at high SFRs the median points show a clear deviation from unity. \citet{2011ApJ...738..106W} are able to find a better match between $SFR_{\rm{SED}}$ and $SFR_{24}$ if they tune key parameters of the SED fit, such as $\tau_{min}$, the $e$-folding time of the exponentially declining star formation history.  The exact tuning needed varies based on several other assumptions in the SED fitting procedure, such as different stellar population synthesis codes, and even when the tuning is done, the computed $SFR_{\rm{SED}}$ still systematically underestimates the true SFR for a significant fraction of sources.  

\begin{figure*}[!h]
\centering
\includegraphics[width=12cm]{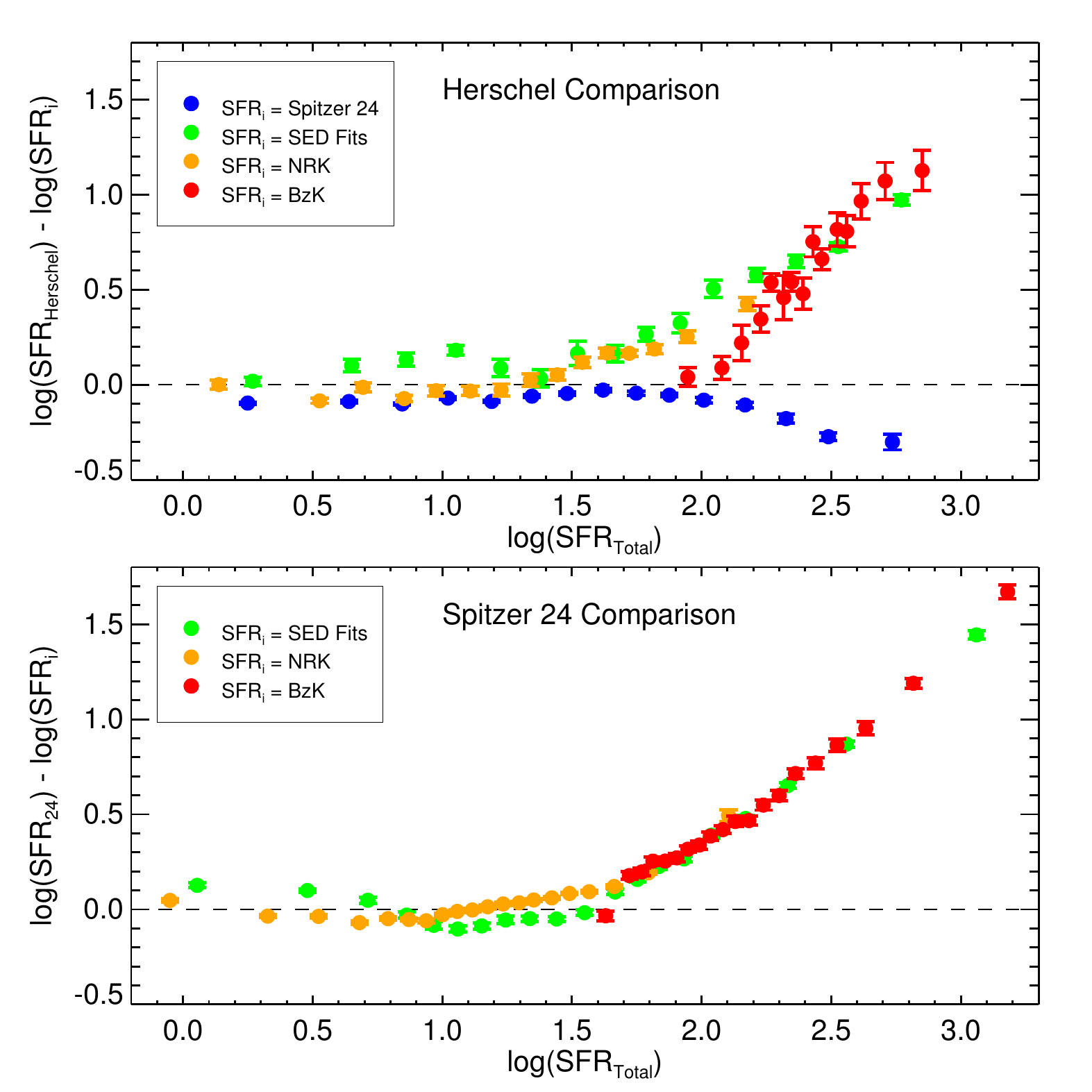}
\caption{Median difference in SFR estimated from infrared vs. optical/UV indicators, as a function of Total SFR. Sources that were detected by both SFR indicators were split in 15 equally populated $SFR_{Total}$ bins, and the median value is plotted, with error bars representing bootstrapped errors. We see that all indicators begin to deviate significantly at $log(SFR) \gtrsim 1.5$. {\bf (Top)} Comparison of {\em Spitzer} 24 $\mu$m + GALEX (blue), multi-wavelength SED fits (green), NRK (orange, $0 < z < 1.3$), and BzK (red, $1.4 < z < 2.5$) to total SFR as derived from {\em Herschel}. {\bf (Bottom)} Comparison of multi-wavelength SED fits (green), NRK (orange, $0 < z < 1.3$), and BzK (red, $1.4 < z < 2.5$) to total SFR as derived from {\em Spitzer} 24 $\mu$m.  }
\label{fig:sfr_comp_diff}
\end{figure*}

Finally, the comparison between $SFR_{\rm{BzK}}$ and $SFR_{\rm{Total}}$ shows essentially no correlation ($\rho_{BzK} = -0.11$).  It should be noted that the redshift range of the BzK indicator ($1.4 < z < 2.5$) limits us to a small sample size containing only the brightest galaxies. As seen in \autoref{fig:sfr_comp_diff}, this selection limits our comparison to galaxies at SFRs where all indicators begin to deviate significantly from $SFR_{\rm{Total}}$. Stacking analyses suggest a stronger correlation between average $SFR_{\rm{BzK}}$ and average $SFR_{Herschel}$ at fainter luminosities \citep{2014MNRAS.443...19R}, but the tightness of the distribution is not well determined. The BzK galaxies that are {\em Herschel} detected show no correlation between the SFR derived from the BzK method and from {\em Herschel} measurements. 

\subsubsection{Selection Effects of SFR Indicators}

The comparisons of the various SFR indicators shown in \autoref{fig:sfr_comp} span different dynamic ranges in SFR, mostly due to the redshift limitations of the NRK and BzK indicators. In \autoref{fig:sfr_comp_diff}, we plot the typical difference between SFR indicators and $SFR_{\rm{Total}}$ as a function of $SFR_{\rm{Total}}$. We see that all of the SFR indicators provide poor estimates of the infrared measured $SFR_{\rm{Total}}$ above $log(SFR) \gtrsim 1.5$ ($\sim 30 M_{\odot}/$yr). These common SFR indicators fail to accurately estimate the true SFR of luminous infrared galaxies. This highlights the need for direct infrared observations to accurately measure the SFR of highly star forming galaxies.  

Throughout this analysis we have assumed the {\em Herschel}-determined SFR is the most accurate because it directly probes far-infrared wavelengths, where the bulk of the re-radiated radiation from dust is emitted. However, it is possible that the high detection threshold of {\em Herschel} limits us to a biased sample that does not accurately reflect the emission properties of lower luminosity galaxies. To test this possibility, we re-run our analyses using the much deeper sample of {\em Spitzer} 24 $\mu$m-detected galaxies (which showed excellent agreement with {\em Herschel} SFRs) as the comparison sample. We find very similar results as the {\em Herschel} comparison, with NRK providing both the strongest correlation and the tightest distribution. 

\subsubsection{A Ladder of SFR Indicators}

While all three non-infrared based SFR indicators fail to accurately estimate the SFR in high luminosity galaxies, the NRK method provides the most accurate and consistent estimates across the full dynamical range of {\em Herschel} SFRs. At high SFRs ($SFR \gtrsim 30 M_{\odot}$/yr), 70\% of our sample is directly detected in the infrared by either {\em Herschel} or {\em Spitzer}.  Thus, we can study the full population of star forming galaxies by constructing a ``ladder'' of SFR indicators \citep[as in][]{2011ApJ...738..106W} based on the {\em Herschel}, {\em Spitzer} 24 $\mu$m, and NRK SFR indicators.  All sources have $SFR_{\rm{Total}}$ calculated using Equation 1, with different methods of determining $L_{\rm{IR}}$.  For sources detected by {\em Herschel}, we measure $\lir$ from fitting the far-infrared photometry to the \citet{2012MNRAS.425.3094C} greybody plus power-law models. We use the $\lir$ estimated from 24 $\mu$m \citep{2009ApJ...703..222L} for sources that are not detected by {\em Herschel} but are detected at {\em Spitzer} 24 $\mu$m. And for the remaining sources, we estimate the $\lir$ using the NRK-derived IRX (as discussed in \autoref{sec:opt_SFR}). Although we include NRK-derived IRX for all galaxies above our mass-completeness limits, the method has only been well-calibrated for $M_{*} > 10^{9.3} M_{\odot}$. Infrared stacking suggests that any systematic offsets should be small, but when calculating main sequence relationships we only include galaxies with $M_{*} > 10^{9.3} M_{\odot}$.

\begin{figure*}[!h]
\centering
\includegraphics[width=14 cm]{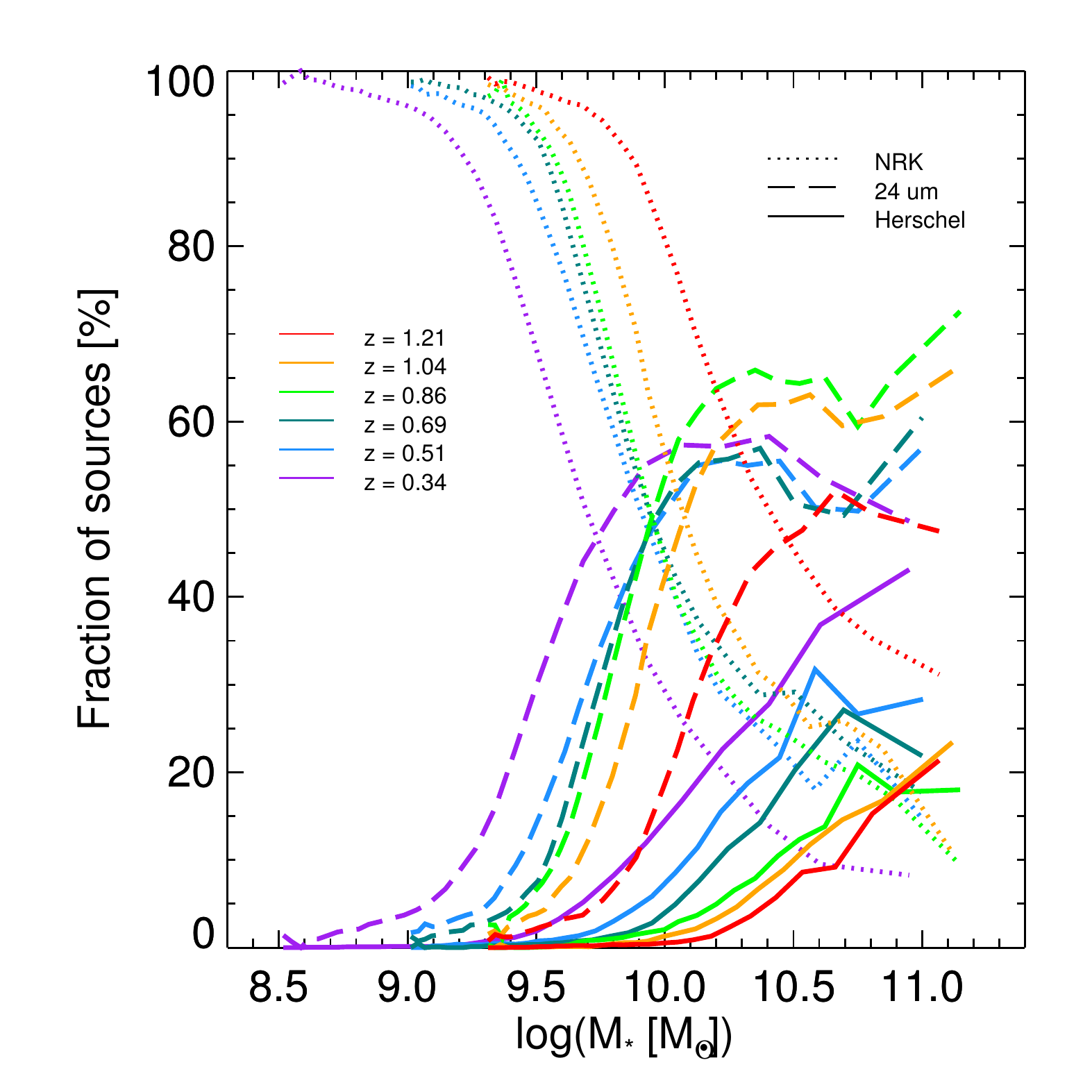}
\caption{Relative fraction of sources with SFRs determined from NRK (dotted line), {\em Spitzer} 24 $\mu$m (dashed line), and {\em Herschel} (solid line). Different colored lines represent percentages in different redshift bins, with bluer colors representing low redshifts and redder colors representing high redshifts.}
\label{fig:nrk_frac}
\end{figure*}

The relative fraction of sources with SFRs measured from each indicator is plotted in \autoref{fig:nrk_frac} as a function of both stellar mass and redshift. Below $M_{*} \lesssim 10^{9.5} M_{\odot}$, SFRs are almost all determined from NRK, but at higher stellar masses, the fraction of sources with direct infrared measurements increases until about 25\% (60\%) of sources at $M \gtrsim 10^{10.5} M_{\odot}$ have SFRs determined from {\em Herschel} ({\em Spitzer} 24 $\mu$m). Because of the redshift limitations of the NRK method \citep[see][]{2013A&A...558A..67A} and the larger errors associated with the $SFR_{\rm{SED}}$ and $SFR_{\rm{BzK}}$ indicators, we restrict the rest of our analysis to redshifts $0 < z < 1.3$, where we can more accurately measure the SFR of our full sample.

\section{Shape of the Main Sequence of Star Formation}\label{sec:ms}

\begin{figure*}[!h]
\centering
\includegraphics[width=14 cm]{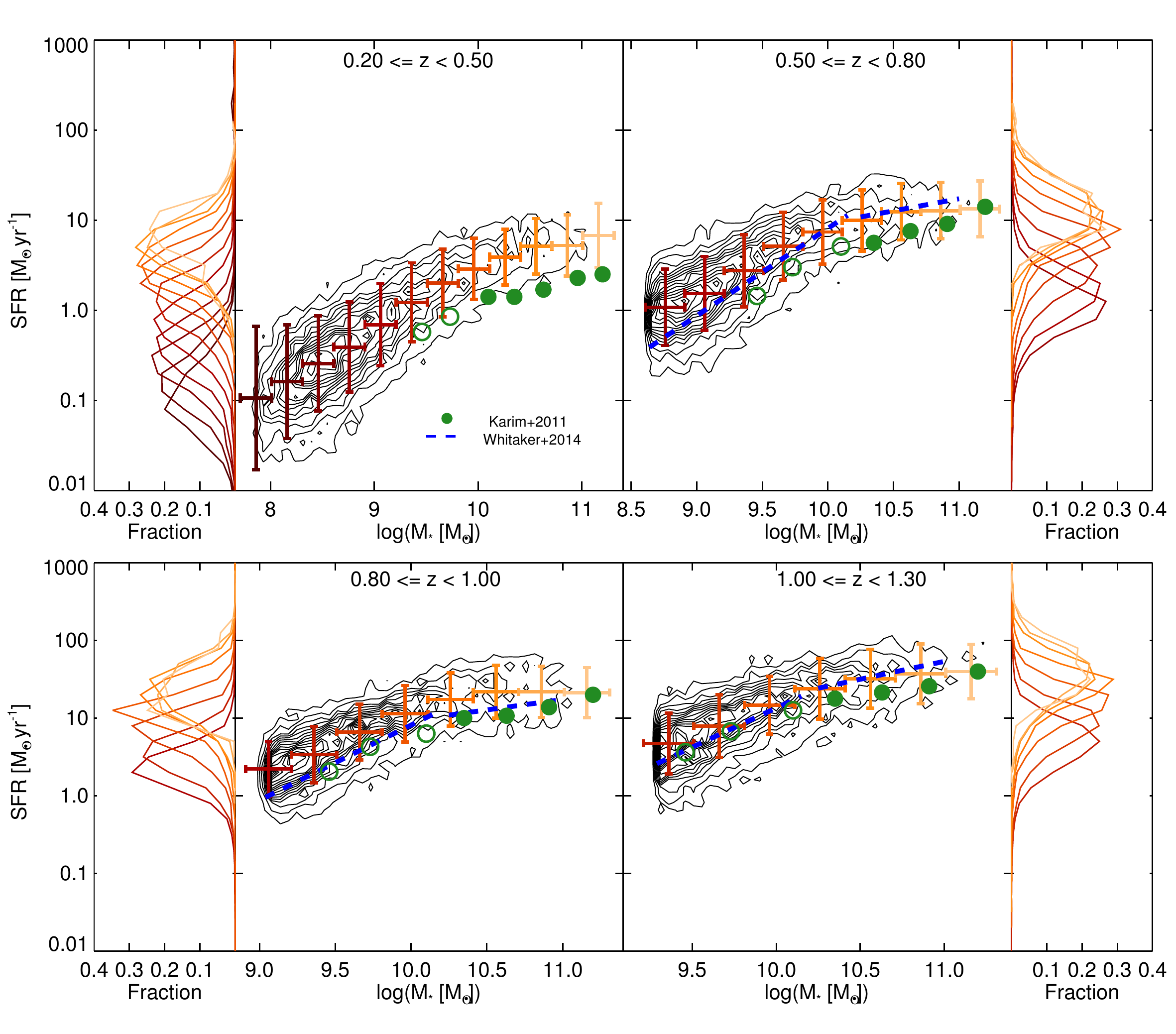}
\caption{Contour density plot of star-forming galaxies in the COSMOS field.  We remove all galaxies classified as ``quiescent'' (unless they were detected in the infrared) and combine all star-forming galaxies, regardless of the specific SFR indicator used.  To display the density of sources in the $SFR/M_{*}$ plane, each redshift slice was made into a grid of $51 \times 51$ bins, and the number of sources in each bin was calculated.  Black contours show the density of galaxies, with contour levels set at 1/2 of the standard deviation of the number of sources in each bin. Colored vertical bars represent the median SFR in mass bins of width 0.3 dex and display the overall trend of the $SFR/M_{*}$ relationship.  Histograms of matching color display the distribution of SFR in each mass bin along the sides of each plot. Main-sequence relationships from \citet[][green dots]{2011ApJ...730...61K} and \citet[][blue line]{2014ApJ...795..104W} are plotted for comparison.}
\label{fig:nrk_mseq_hist}
\end{figure*}

With reliable and consistent SFR estimates for a large, mass-complete sample of galaxies in COSMOS, we examine the star-forming main sequence for a large, unbiased sample of 62,521 galaxies. \autoref{fig:nrk_mseq_hist} displays the stellar mass and SFR of our full sample, split into four redshift bins spanning $0.2 \le z \le 1.3$. Black contours display the density of sources at each location in SFR and $M_{*}$ parameter space, and colored bars represent the median SFR in stellar mass bins of width $\Delta$log($M_{*}$) = 0.3, with vertical error bars displaying the standard deviation of the SFRs in that bin. These same bins are used to create the fractional histograms plotted on the side of each redshift bin, which display the distribution of SFRs within each mass bin with the corresponding color. The derived main sequence relationships from star-forming galaxies in \citet{2011ApJ...730...61K} and \citet{2014ApJ...795..104W} are also plotted for comparison. 

\autoref{fig:nrk_mseq_hist} shows that the galaxies in our sample do not follow a simple linear main sequence relationship between $log(SFR)$ and $log(M_{*})$ (or a single power-law relationship between $SFR$ and $M_{*}$).  Instead, the median SFR relationship appears to flatten at masses above $M_{*} \sim 10^{10} M_{\odot}$.  This can be seen in the histograms, which show that the peak of each SFR distribution increases with increasing $M_{*}$ at low masses, but at high masses, the histogram peaks all lie at approximately the same SFR. The standard deviation of the SFR in each stellar mass bin remains mostly constant at all masses and at all redshifts, with $\sigma \sim 0.36$ dex in all bins. The shape of this relationship appears roughly constant with redshift, with the entire relationship increasing to higher SFRs at higher redshifts. 

\subsection{Parameterizing the Star-Forming Main Sequence}\label{sec:ms_fits}

\begin{figure*}[!h]
\centering
\includegraphics[width=14 cm]{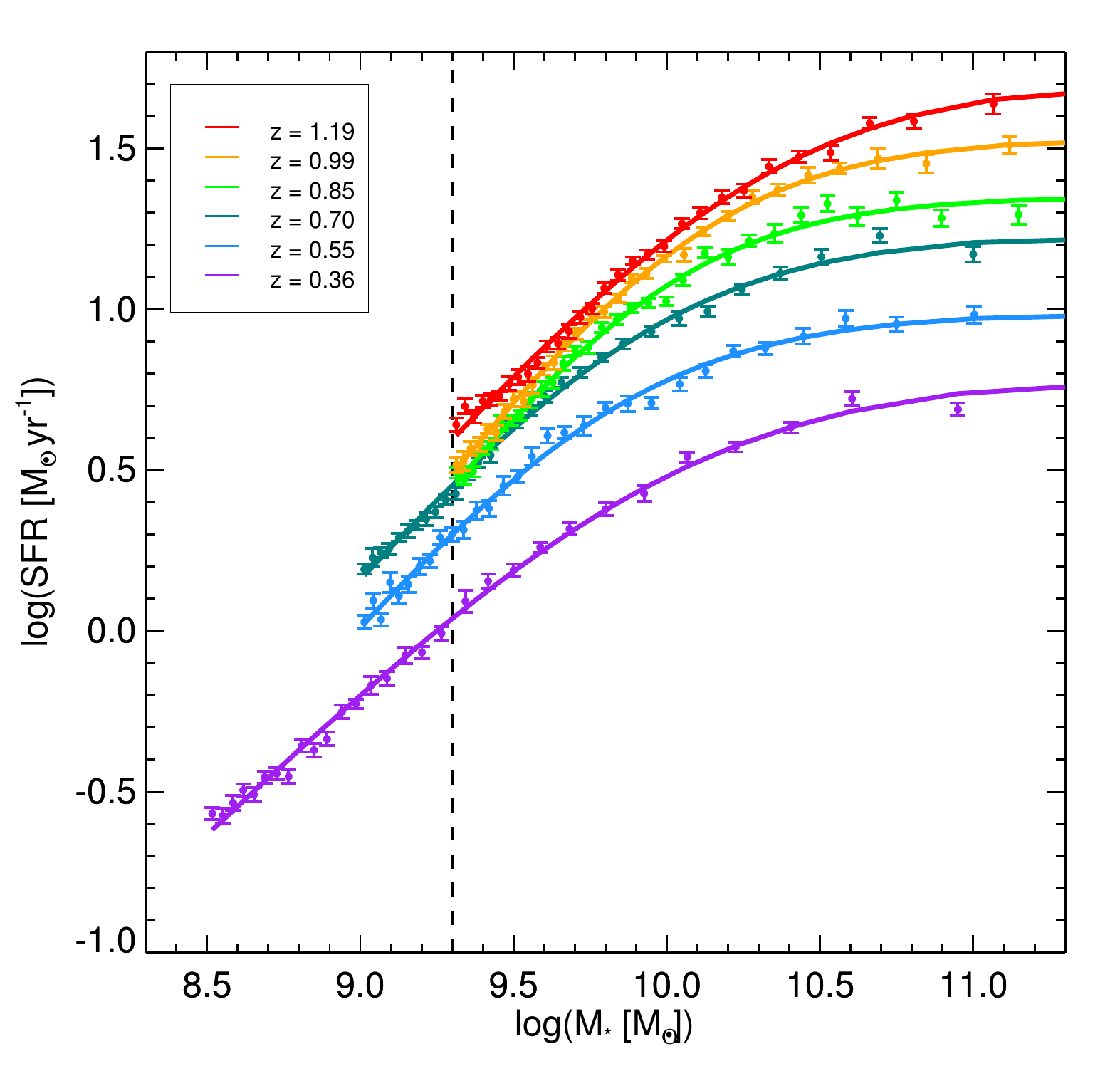}
\caption{Median SFR in 6 equally populated redshift bins that have been split into 30 equally populated stellar mass bins. Errors on the median are calculated from bootstrapping. Solid lines represent the best-fit curve to the model $\mathcal{S} = \mathcal{S}_{0} - log \left[ 1 + \left( \frac{10^{\mathcal{M}}}{10^{\mathcal{M}_{0}}} \right)^{- \gamma} \right]$. Vertical dashed line represents the stellar mass limit below which NRK has not been well-calibrated.}
\label{fig:nrk_mseq_allz}
\end{figure*}

From \autoref{fig:nrk_mseq_hist}, it is clear that a single power-law does not accurately describe the relationship between stellar mass and star formation rate. We split our sample of star-forming galaxies into 6 equally populated redshift bins, each of which are then split into 30 equally populated stellar mass bins (with $\sim 350$ sources in each bin), and calculate the median SFR in each bin. We limit our sample to stellar masses above a conservative mass limit (see \autoref{tab:param}) to ensure that we are not affected by systematics. The specific number of redshift bins does not affect the following results, although we must balance between having redshift bins that are too wide and combine galaxy samples at different epochs with having redshift bins that are too narrow and are affected by small number statistics. The same is true for the number of stellar mass bins, although having at least 30 bins is preferable for accurately determining the goodness of fit to the models.  The median SFRs for every bin are plotted in \autoref{fig:nrk_mseq_allz}, colored by redshift, with bootstrapped errors on the median represented by vertical bars.

Using the {\em MPFIT} package implemented in IDL \citep{2009ASPC..411..251M}, we fit the median $log(M_{*})$ and $log(SFR)$ in each redshift bin with many models including linear, $2^{\rm{nd}}$ order polynomial, and broken linear, and find the best fit is provided by the following model:
\begin{equation}\label{eqn:ms_fit}
\mathcal{S} = \mathcal{S}_{0} - log \left[ 1 + \left( \frac{10^{\mathcal{M}}}{10^{\mathcal{M}_{0}}} \right)^{- \gamma} \right]
%\mathcal{S} = \mathcal{S}_{0} + log \left[ 1 - \exp\Bigg(-\Big(\frac{10^{\mathcal{M}}}{10^{\mathcal{M}_{0}}}\Big)^{\gamma}\Bigg) \right]
\end{equation}
where $\mathcal{S} = log(SFR)$ and $\mathcal{M} = log(M_{*}/M_{\odot})$. We choose this model because (1) at all redshifts, it provides the best reduced $\chi^{2}$ fit to the data, and  (2) unlike polynomial fits, the parameters of the model allow us to quantify the interesting characteristics of the relation between stellar mass and SFR: $\gamma$, the power-law slope at low stellar masses, $\mathcal{M}_{0}$, the turnover mass (in $log(M_{*}/M_{\odot})$), and $\mathcal{S}_{0}$, the maximum value of $\mathcal{S}$ (or the maximum value of $log(SFR)$) that the function asymptotically approaches at high stellar mass. The best-fit parameters for each redshift bin are listed in Table \ref{tab:param}. 

\begin{comment}
While we must use caution to not over-interpret the parameters of our model, it provides us with a framework within which to analyze our data (unlike a 2nd order polynomial, for example). At all redshifts, this logarithmic model fit the data better than any of the other models we tried, although this does not mean that it is the only possible interpretation of the data. With these caveats in mind, we examine how these parameters evolve with redshift, and what this means for star-formation and stellar mass buildup in the universe. 
\end{comment}

\subsection{Evolution of Model Parameters}\label{sec:params}

\begin{figure*}[!h]
\centering
\includegraphics[width=16 cm]{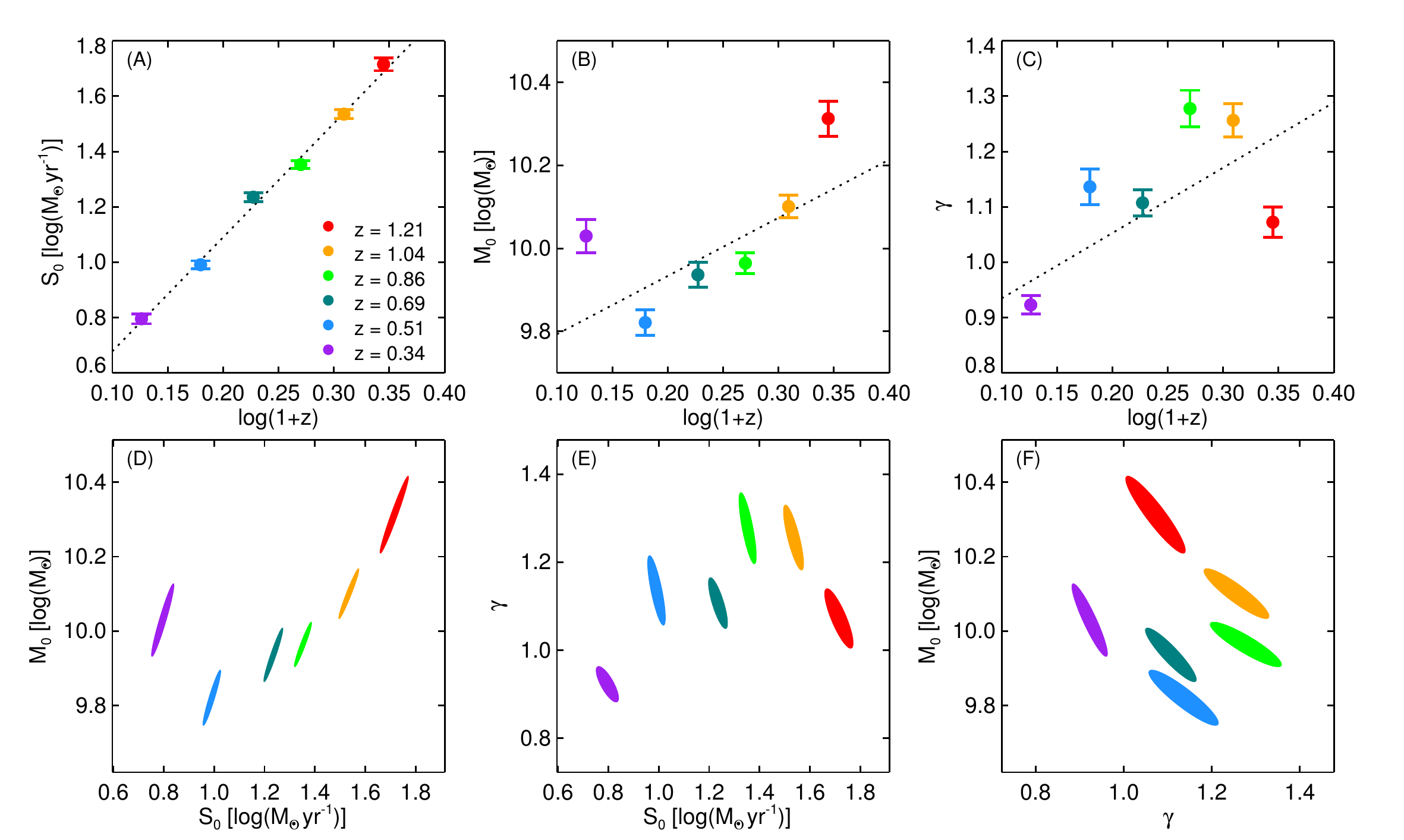}
\caption{{\bf Top:} Redshift evolution of the best-fit parameter values for {\em (A)} maximum of $log(SFR)$, $\mathcal{S}_{0}$; {\em (B)} turnover mass, $\mathcal{M}_{0}$; and {\em (C)} power-law slope, $\gamma$. Different color dots represent parameter values in different redshift bins with 1-$\sigma$ error bars, and the dotted line is the best-fit linear fit to the data.  {\bf Bottom:} 95\% confidence error ellipses displaying the covariance between {\em (D)} $\mathcal{S}_{0}$ and $\mathcal{M}_{0}$; {\em (E)} $\mathcal{S}_{0}$ and $\gamma$; and {\em (F)} $\gamma$ and $\mathcal{M}_{0}$, with different colors once again representing different redshift bins. We see moderate covariance in $\gamma$ and $\mathcal{M}_{0}$, but little covariance between the other pairs. }
\label{fig:nrk_mseq_param}
\end{figure*}

In the top panels of \autoref{fig:nrk_mseq_param}, we plot the evolution of $\mathcal{S}_{0}$, $\mathcal{M}_{0}$, and $\gamma$ as functions of $log(1+z)$. The bottom panels of \autoref{fig:nrk_mseq_param} examine the covariance between these parameters by displaying the 95\% confidence error ellipses. 

We see clear and strong evolution of $\mathcal{S}_{0}$ with redshift, and the best fit line suggests an evolution of $\mathcal{S}_{0} \propto (4.12 \pm 0.10) \times log(1+z)$, or equivalently, $SFR_{0} \propto (1+z)^{4.12 \pm 0.10}$. The covariance between $\mathcal{S}_{0}$ and $\mathcal{M}_{0}$ (\autoref{fig:nrk_mseq_param}D) and between $\mathcal{S}_{0}$ and $\gamma$ (\autoref{fig:nrk_mseq_param}E) is relatively minor, so we infer that the evolution in $\mathcal{S}_{0}$ is true evolution and not due to variation in the other parameters. 

Both $\mathcal{M}_{0}$ and $\gamma$ show some evidence for weak evolution to more massive $\mathcal{M}_{0}$ and steeper $\gamma$ with redshift, although much of the perceived evolution may be due to the covariance seen in \autoref{fig:nrk_mseq_param}F.  The best (linear) fit to the evolution in the turnover mass is given by $\mathcal{M}_{0} \propto (1.41 \pm 0.20) \times log(1+z)$. We test the possible redshift evolution of turnover mass by calculating where the data deviates by 0.2 dex from a single power-law fit to the low mass data and find similar evolution, suggesting that the turnover mass does indeed change with cosmic time.  The low-mass power-law slope, $\gamma$, has a best-fit line that suggests evolution of $\gamma \propto (1.17 \pm 0.13) \times log(1+z)$. Redshift evolution in $\gamma$ to steeper slopes at earlier cosmic times would suggest that the SFR in the lowest stellar mass galaxies does not increase as much as in more massive systems.

\subsection{Separating Quiescent Galaxies}\label{sec:quiescent}

We have described the main sequence relationship between $SFR$ and $M_{*}$ for {\em star-forming} galaxies. However, possible misclassification of galaxies as either ``star-forming'' or ``quiescent'' could drastically affect the trends we observe. 

\begin{figure*}[!h]
\centering
\includegraphics[width=14 cm]{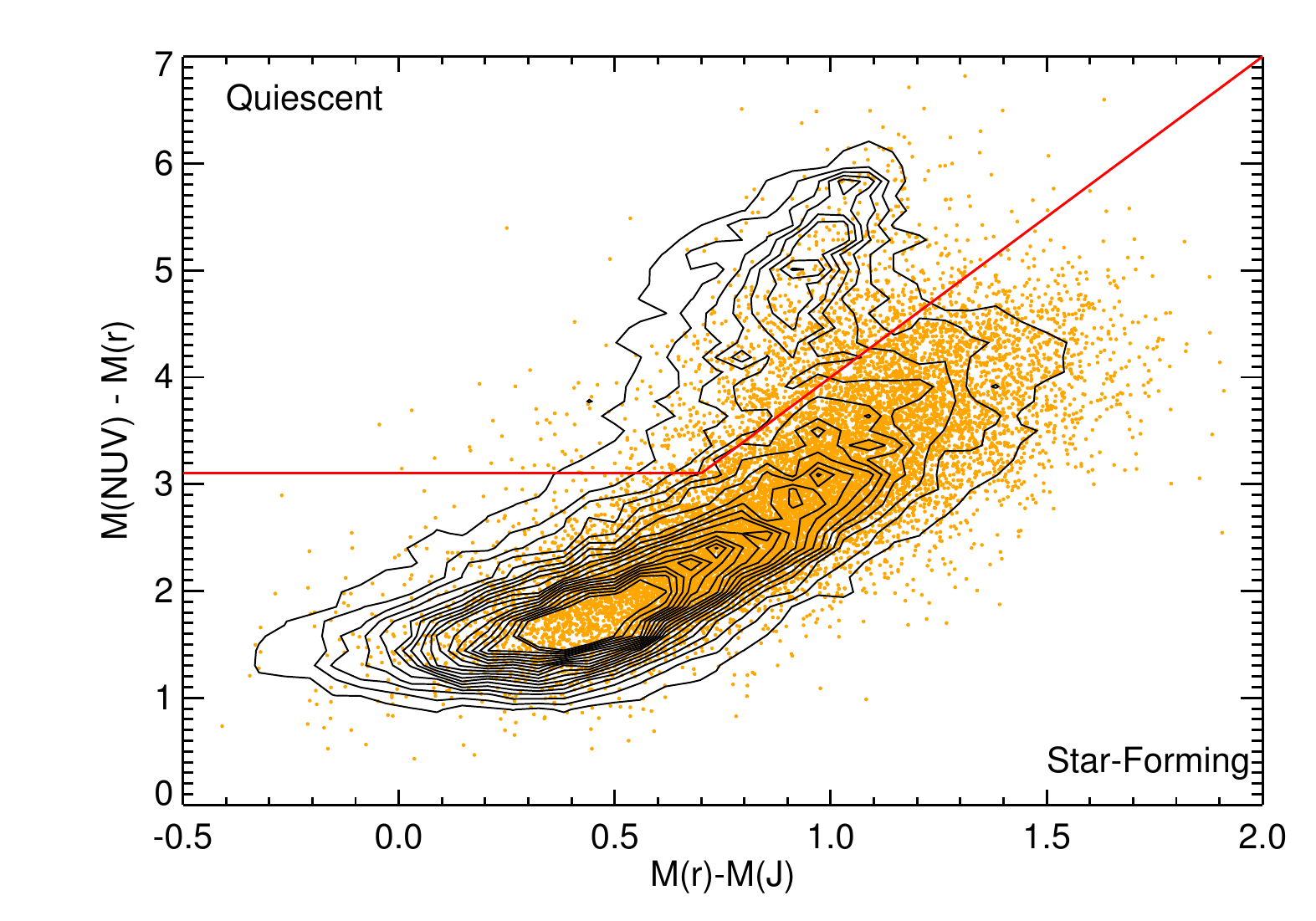}
\caption{$NUV - r^{+}$ vs. $r^{+} - J$ plot of a mass-selected sample of galaxies in COSMOS at $0.2 < z < 1.3$. Black contours represent the full sample of galaxies, while orange circles highlight galaxies that are IR-detected, either with {\em Herschel} or {\em Spitzer} 24 $\mu$m.  The red line, from I13, divides the sample into ``star-forming'' and ``quiescent'' galaxies.  The majority of IR-detected galaxies are properly classified as ``star-forming'', but there is a significant population ($\sim 7$\%) that are misclassified as ``quiescent''.}
\label{fig:color_mag}
\end{figure*}

As described in \autoref{sec:ms}, we remove galaxies that are considered quiescent from our sample using the selection $M_{\rm{NUV}} - M_{r} > 3(M_{r} - M_{J}) + 1$ and $M_{NUV} - M_{r} > 3.1$ \citep{2013A&A...556A..55I}. This selection is shown in \autoref{fig:color_mag}, with the full mass-selected sample of COSMOS galaxies at $0.2 < z < 1.3$ generally separated into two distinct ``star-forming'' and ``quiescent'' regions.  Improper classification of the ``in-between'' galaxies that are not obviously star-forming or quiescent could lead to changes in the main sequence shape.  To test this possibility, we shift the entire separating line (both horizontal and diagonal segments) between quiescent and star-forming galaxies by $\pm 0.4$ mag in $M_{\rm{NUV}} - M_{r}$, and in either case there is no appreciable change to the main-sequence. 

Galaxies detected in the infrared by {\em Herschel} or {\em Spitzer} 24 $\mu$m are highlighted in \autoref{fig:color_mag}, and while the majority fall on the star-forming sequence, we see a number of objects that lie in the quiescent region. This population of infrared-detected quiescent (IR-Q) galaxies is relatively small, with only $\sim 7$\% of the galaxies classified as quiescent having detectable infrared emission, but these misclassified galaxies are predominantly found at high stellar mass.  The fraction of quiescent galaxies detected in the infrared increases rapidly from $\le 1$\% at $M_{*} \sim 10^{9.5} M_{\odot}$ to 15-20\% at $M_{*} \ge 10^{11} M_{\odot}$, and this trend holds at all redshifts. These massive galaxies could heavily influence the shape of the main-sequence we observe, so it is vital to understand what is driving their infrared emission. 

There could be several reasons why galaxies with quiescent colors have significant emission in the infrared, including (i) improper classification of star-forming galaxies possibly due to extreme dust obscuration, (ii) elevated infrared luminosity from an AGN, (iii) inaccurate absolute magnitudes due to catastrophic failures in photo-z's or low signal-to-noise photometry, or (iv) ``post-starburst'' infrared glow due to dust heating from young stars (that is not related to the instantaneous star formation). 

AGN typically heat dust to very hot temperatures, so we expect any AGN contribution to infrared radiation to be predominantly in near- and mid-IR wavelengths, while far-infrared emission is likely due to star formation alone. Only about 10\% of the IR-Q galaxies have been detected by {\em Herschel}, and the rest are 24 $\mu$m-only detections, where AGN may heavily influence the emission. However, only $\sim 1$--5\% of the IR-Q galaxies are detected in the X-ray by {\em Chandra}, and only $\sim 2$--8\% of the IR-Q galaxies have IRAC power-law colors indicative of AGN, with significant overlap in those two populations, and the percentages are even lower when looking only at the 24 $\mu$m-only sources. This suggests that radiation from an AGN is not fueling the infrared emission. The average SFR of the IR-Q galaxies with AGN is 0.2--0.5 dex higher than the average SFR of all the IR-Q galaxies, so the presence of AGN in these galaxies is likely just a reflection of the well-studied trend that AGN fraction increases with SFR or $\lir$ \citep[e.g.][]{2010ApJ...709..572K}. 

The rather high SFRs of the IR-Q galaxies suggest that they are indeed driven by star-formation, and have been misclassified as quiescent. Man et al. (in prep) stack the infrared emission from quiescent galaxies and find upper limits of $SFR_{\rm{IR}} < 0.1$--1 $M_{\odot}$/yr. The SFRs of the IR-Q galaxies in our sample tend to lie below the main-sequence, but are all at least $\times 2$--3 higher than the upper limit from Man et al., which suggests that they are indeed still actively star-forming. In addition, the infrared emission from IR-Qs is brighter than expected from a ``post-starburst'' glow \citep{2014arXiv1402.0006H}. It is unlikely that catastrophic photo-z errors or low signal-to-noise photometry are causing these misidentifications, as the sources have excellent photometry and well-constrained photometric redshifts (only $0.1$\% of the IR-Q galaxies have $\sigma_{\Delta z/(1+z)} > 0.15$). Thus, the likely explanation for these sources is that they are actively star-forming galaxies that have been misclassified as ``quiescent'', and we include them in our analysis of the main-sequence. We note, however, that the shape of the main-sequence does not change significantly based on the inclusion or exclusion of these sources.

%\subsection{Systematics that affect the NRK}\label{sec:nrk_err}

\section{Discussion}\label{sec:discussion}

We see that the slope of the main sequence relationship between $SFR$ and $M_{*}$ changes with stellar mass. While most previous studies found a constant main sequence slope \citep[for a summary see][]{2014arXiv1405.2041S}, some recent studies found a curved relationship might provide a better fit to the data \citep{2011ApJ...730...61K,2012ApJ...754L..29W}.  However, the mass completeness limit in both studies coincided with the turnover mass, leaving doubt as to whether the turnover was a real trend or an artifact of completeness.  With our new COSMOS observations, we are able to study star-forming galaxies considerably less massive than the turnover mass, and we find that a single power-law does not provide the best description of the star-forming main-sequence. 

\subsection{The Turnover in the Star-Forming Main Sequence}

The relationship between SFR and $M_{*}$ varies with stellar mass, with two distinct regions below and above the characteristic turnover mass, $\mathcal{M}_{0}$. What causes this change, and why does the turnover occur at about $M_{*} \approx 10^{10} M_{\odot}$ at all redshifts? 

\subsubsection{The slope of the main-sequence}

The parameterization of the main sequence we employ in \autoref{sec:ms_fits} includes a parameter $\gamma$ that we describe as the low stellar mass power-law slope. However, we note that this slope is not derived from an actual power-law fit to the data, but instead represents the power-law slope that the relationship approaches in the very low-mass regime, based on \autoref{eqn:ms_fit}. This slope is significantly steeper than power-law slopes commonly quoted in the literature, and should not be compared to slopes from power-law fits to data. 

For an easier comparison to the existing literature, we derive best-fit power-law relationships, fitting the low mass regime and high mass regime separately. Galaxies less massive than the turnover mass follow a fairly tight power-law relationship of $SFR \propto M^{\beta}$, with $\beta = 0.88 \pm 0.06$. This slope is shallower than $\gamma$ because it includes galaxies in the ``turnover region'', where the slope is already starting to flatten. Galaxies more massive than the turnover mass follow a drastically different relationship, with $\beta = 0.27 \pm 0.04$ for $M_{*} > 10^{10} M_{\odot}$.  A galaxy's specific star formation rate ($SSFR \equiv SFR/M_{*}$) can be interpreted as a measure of the efficiency of current star formation as compared to its past average star formation history. The $SSFR$ of massive galaxies is systematically lower than would be expected from an extrapolation of low mass galaxies, suggesting that there may be decreased star formation efficiency in high stellar mass galaxies. 

\subsubsection{Quenching in High Mass Galaxies}

The turnover in the main sequence to lower star formation efficiencies in massive galaxies suggests there is a fundamental change that occurs as galaxies become more massive, as has been predicted in some studies.  Galaxy luminosity and mass functions, which measure the brightness and mass distribution of galaxies at various lookback times, show a steep, exponential decline at high stellar masses and high luminosities while retaining a remarkably consistent shape at all redshifts \citep[e.g.][]{2003ApJS..149..289B,2010A&A...523A..13P,2013A&A...556A..55I}. The lack of large, bright galaxies throughout cosmic time argues for the presence of a characteristic mass above which a galaxy is likely to have its star formation strongly suppressed or quenched.  

In contrast, the {\it dark} matter halo mass function from semi-analytic models does not show the same exponential decline, and instead has a much shallower power-law cutoff at much higher masses \citep{1999MNRAS.310.1087S,2003ApJ...599...38B}, leading to a ``pivot mass'' above which the ratio of dark matter to light matter increases rapidly \citep[e.g.][]{2012ApJ...744..159L}. At low stellar masses, the stellar to halo mass relation \citep[$M_{h} \propto M_{*}^{0.46}$;][]{2012ApJ...744..159L} and the dark matter halo growth rates from $N$-body simulations \citep[$\dot{M}_{halo} \propto (M_{halo})^{1.1}$;][]{2002ApJ...568...52W,2009MNRAS.398.1858M,2010MNRAS.401.2245F} suggest a main sequence relationship of $SFR \propto M_{*}^{1.04}$, similar to the slope seen in the main-sequence.  The ``pivot mass'', above which the stellar-to-halo mass relation deviates from the low stellar mass relationship, appears to evolve to higher $M_{*}$ at higher redshifts  \citep{2012ApJ...744..159L}, at a rate similar to the possible evolution seen in the main sequence turnover mass, $\mathcal{M}_{0}$. In galaxies more massive than the ``pivot mass,'' the halo mass rises sharply in comparison with stellar mass, suggesting that while massive dark matter haloes appear to continue growing, the galaxies residing in them quench their star formation.  

Possible mechanisms for this quenching include structural disruptions or galaxy mergers \citep{1996ARA&A..34..749S,2006ApJS..163....1H}, feedback from accretion onto a supermassive black hole \citep{2005ApJ...620L..79S}, gravitational heating of the surrounding intracluster medium \citep{2008ApJ...680...54K}, changes in the mode of gas accretion onto galaxies \citep{2005MNRAS.363....2K,2007MNRAS.380..339B,2013MNRAS.429.3353N}, or gas removal or strangulation in dense environments \citep{2010ApJ...721..193P,2012ApJ...757....4P}.  

Morphological studies may be key for understanding the star-formation in massive galaxies. \citet{2014ApJ...785L..36A} find that galaxy SFRs are more strongly correlated to disk stellar mass (as opposed to total stellar mass), and that $SSFR_{disk}$ is approximately constant with mass.  If this is the case, the turnover in the main-sequence could be simply due to growing bulges in the highest mass systems. However, one might expect the turnover to disappear (or become less severe) at high redshifts as galaxies become more disk-dominated, but this is not seen in the data. \citet{2014MNRAS.tmp..527S} find that disk galaxies and elliptical galaxies likely quench their star formation rates through different processes with very different timescales. A galaxy's physical size may also play a role in quenching, as the surface mass density has been shown to correlate strongly with $SSFR$ \citep{2003MNRAS.341...54K,2008ApJ...688..770F}, and compact star forming galaxies may be on the evolutionary path toward quiescent galaxies \citep{2014ApJ...791...52B}. 

Our data suggest that galaxies with high stellar mass ($M_{*} > 10^{10} M_{\odot}$) are forming stars at a lower rate than would be expected from extrapolating the trends of low stellar mass galaxies. Finding the possible causes of this ``quenching'' of star formation is one of the key hurdles for understanding galaxy evolution. The existence of a ``turnover mass'' hint that the stellar mass of a galaxy plays a crucial role in quenching, possibly related to the ``mass quenching'' discussed in \citet{2010ApJ...721..193P}. Further study is needed to determine the physical mechanism(s) behind quenching.

\subsection{Increasing SFR with Redshift}

From our fits, we find strong evolution in $\mathcal{S}_{0}$, which parameterizes the overall scaling of the $SFR/M_{*}$ main-sequence with redshift.\footnote{Alternatively, measuring the redshift evolution of the SFR at a constant characteristic mass provides similar results} The scaling of the main sequence has been found in the literature to evolve as $(1+z)^{n}$, with the exponent $n$ varying from $2.2 < n < 5$ \citep{2006ApJ...647..128E,2007ApJ...670..156D,2009ApJ...690..937D,2009MNRAS.394....3D,2009ApJ...698L.116P,2011ApJ...730...61K}. The value of $n = 4.12 \pm 0.10$ we measure is among the steeper slopes seen in the literature.  

It's thought that the redshift evolution of the main-sequence normalization is due, at least in part, to increasing gas content in galaxies at earlier cosmic times. However, measuring the gas content in galaxies can be difficult, especially in high redshift systems. Molecular hydrogen is notoriously difficult to detect, so many surveys instead probe the rotational transitions of CO and use locally calibrated CO-to-H$_{2}$ conversion factors, although this conversion factor may differ in starburst galaxies \citep{2008ApJ...680..246T,2011ApJ...740L..15M,2012A&A...548A..22M}. Another method to estimate gas content is to measure the dust mass from far-infrared or submillimeter photometry and convert to gas masses using an assumed gas-to-dust ratio \citep[e.g.][]{2012ApJ...760....6M,2014A&A...562A..30S,2014ApJ...783...84S}. \citet{2012ApJ...760....6M} find gas fraction evolves as $(1+z)^{2.8}$, while recent ALMA observations suggest a steeper evolution of $(1+z)^{5.9}$ \citep{2014ApJ...783...84S}. \citet{2014ApJ...791..130Z} study the mass metallicity relationship and infer a much shallower evolution of gas mass $M_{g} \propto (1+z)^{1.35}$.  Future studies will be needed to determine if the evolving normalization of the main-sequence can be explained simply by an increasing gas supply in galaxies, or if other explanations such as increased merger rates or increased star formation efficiency are necessary to fully explain the observed evolution.

\section{Conclusions}\label{sec:conclusions}

Using new far-infrared data from {\em Herschel}, we compare direct measurements of unobscured and obscured SFR with various SFR indicators that estimate the obscured SFR from data at shorter wavelengths (usually in optical or UV), and find that the NRK method of \citet{2013A&A...558A..67A} provides the most consistent estimate of the far-infrared derived SFR.  By combining the SFRs from {\em Herschel}, {\em Spitzer}, and NRK, we analyze the relationship between $SFR$ and $M_{*}$ (commonly referred to as the ``star-forming main-sequence'') in 62,521 star-forming galaxies at $z \le 1.3$ in the COSMOS field. From our new analysis we find:
\begin{itemize}
\item The relationship between SFR and stellar mass does not follow a simple power-law, but flattens to near-constant SFRs at high stellar masses.  The shape of the main sequence is roughly constant for all redshifts $z \le 1.3$. 
\item The scaling of the entire star-forming main sequence rises with redshift as $(1+z)^{4.12 \pm 0.10}$.
\item The characteristic turnover mass lies at $M_{0} \approx 10^{10} M_{\odot}$, with possible evolution toward higher turnover masses at high redshift.
\item The slope of the low-mass power-law lies between $\gamma = 0.9$--1.3, with possible weak evolution toward steeper slopes at higher redshift.
\item A broken power-law fit to galaxies below and above the turnover mass results in $SFR \propto M_{*}^{0.88 \pm 0.06}$ below the turnover mass and $SFR \propto M_{*}^{0.27 \pm 0.04}$ above the turnover mass. 
\end{itemize}
Our analysis suggests that star-forming galaxies cannot be described by a single power-law relationship between $SFR$ and $M_{*}$, as had been suggested in many previous studies.  Because of the strong effects of dust, direct observations in the FIR are crucial for studying the entire population of star-forming galaxies. In future work we will explore possible causes of the turnover in the main sequence by studying detailed morphology and examining possible feedback mechanisms, and we will extend our analysis to higher redshifts.  

\section{Acknowledgements}
D. B. Sanders and C. M. Casey acknowledge the hospitality of the Aspen Center for Physics, which is supported by the National Science Foundation Grant No. PHY-1066293. C. M. Casey would like to acknowledge generous support from a McCue Fellowship through the University of California, Irvine's Center for Cosmology. KS gratefully acknowledges support from Swiss National Science Foundation Grant PP00P2\_138979/1. KS acknowledges support from the National Radio Astronomy Observatory, which is a facility of the National Science Foundation operated under cooperative agreement by Associated Universities, Inc.

COSMOS is based on observations with the NASA/ESA {\em Hubble Space
Telescope}, obtained at the Space Telescope Science Institute, which
is operated by AURA Inc, under NASA contract NAS 5-26555; also based
on data collected at: the Subaru Telescope, which is operated by the
National Astronomical Observatory of Japan; the XMM-Newton, an ESA
science mission with instruments and contributions directly funded by
ESA Member States and NASA; the European Southern Observatory, Chile;
Kitt Peak National Observatory, Cerro Tololo Inter-American
Observatory, and the National Optical Astronomy Observatory, which are
operated by the Association of Universities for Research in Astronomy,
Inc. (AURA) under cooperative agreement with the National Science
Foundation; the National Radio Astronomy Observatory which is a
facility of the National Science Foundation operated under cooperative
agreement by Associated Universities, Inc; and the
Canada-France-Hawaii Telescope operated by the National Research
Council of Canada, the Centre National de la Recherche Scientifique de
France and the University of Hawaii.

 The Dark Cosmology Centre is funded by the Danish National Research Foundation.

%%%%%%%%%%%
%%% TABLES %%%
%%%%%%%%%%%
\clearpage

\capstartfalse   % Make Hyper Ref work
\begin{deluxetable}{ccccccc}
\tabletypesize{\footnotesize}
\tablewidth{460pt}
%% This is the title of the table.
%\tablecaption{Best Fit Parameters}
\tablehead{\colhead{Redshift Range} & \colhead{$<z>$} & \colhead{$log(M_{limit})$} & \colhead{$\mathcal{S}_{0}$} & \colhead{$\mathcal{M}_{0}$} & \colhead{$\gamma$} & \colhead{Reduced $\chi^{2}$} \\
\colhead{} & \colhead{} & \colhead{$log(M_{\odot})$} & \colhead{$log(M_{\odot}/yr)$} & \colhead{$log(M_{\odot})$} & \colhead{} & \colhead{}} 
%% All data must appear between the \startdata and \enddata commands
\startdata
0.25--0.46 & 0.36 & 8.50 & 0.80 $\pm$ 0.019 & 10.03 $\pm$ 0.042 & 0.92 $\pm$ 0.017 & 1.74 \\
0.46--0.63 & 0.55 & 9.00 & 0.99 $\pm$ 0.015 & 9.82 $\pm$ 0.031 & 1.13 $\pm$ 0.033 & 1.52 \\
0.63--0.78 & 0.70 & 9.00 & 1.23 $\pm$ 0.016 & 9.93 $\pm$ 0.031 & 1.11 $\pm$ 0.025 & 1.48 \\
0.78--0.93 & 0.85 & 9.30 & 1.35 $\pm$ 0.014 & 9.96 $\pm$ 0.025 & 1.28 $\pm$ 0.034 & 1.84 \\
0.93--1.11 & 0.99 & 9.30 & 1.53 $\pm$ 0.017 & 10.10 $\pm$ 0.029 & 1.26 $\pm$ 0.032 & 0.62 \\
1.11--1.30 & 1.19 & 9.30 & 1.72 $\pm$ 0.024 & 10.31 $\pm$ 0.043 & 1.07 $\pm$ 0.028 & 1.24 \\
\enddata
\label{tab:param}
%% General table comment marker
\tablecomments{Parameters of the best fit model to the star-forming main sequence. The full sample of 62,521 star-forming galaxies is split into six equally populated bins, with each bin containing $\sim 17745$ galaxies. Within each redshift bin, the galaxies are split into 30 equally populated bins of stellar mass.  The median SFR in each mass bin is calculated and then fit to $\mathcal{S} = \mathcal{S}_{0} - log \left[ 1 + \left( \frac{10^{\mathcal{M}}}{10^{\mathcal{M}_{0}}} \right)^{- \gamma} \right]$, where $\mathcal{S} = log(SFR)$ and $\mathcal{M} = log(M_{*})$. Table columns are as follows: {\bf(1)} Redshift range of bin; {\bf(2)} Median Redshift; {\bf(3)} Stellar Mass Limit of redshift bin; {\bf(4)} $\mathcal{S}_{0}$, the maximum value of $\mathcal{S}$; {\bf(5)} Turnover Mass; {\bf(6)} Low-mass power-law slope; {\bf(7)} Reduced $\chi^{2}$ of fit.}
\end{deluxetable}
\capstarttrue  % Make Hyper Ref work

\clearpage
\bibliographystyle{apj}
\bibliography{apj-jour,main_sequence}

\end{document}